# Systematic Analysis of COVID-19 Ontologies

Debanjali Bain[1] [0000–0002–2811–6692] and Biswanath Dutta[2] [0000–0003–3059–8202]

[1, 2] Documentation Research and Training Centre (DRTC)
Indian Statistical Institute, Bangalore, India
[1] debanjali@drtc.isibang.ac.in
[1] Department of Library and Information Science, Calcutta University, Kolkata, India
[2] bisu@drtc.isibang.ac.in

**Abstract**. This comprehensive study conducts an in-depth analysis of existing COVID-19 ontologies, scrutinizing their objectives, classifications, design methodologies, and domain focal points. The study is conducted through a dual-stage approach, commencing with a systematic review of relevant literature and followed by an ontological assessment utilizing a parametric methodology. Through this meticulous process, twenty-four COVID-19 Ontologies (CovOs) are selected and examined. The findings highlight the scope, intended purpose, granularity of ontology, modularity, formalism, vocabulary reuse, and extent of domain coverage. The analysis reveals varying levels of formality in ontology development, a prevalent preference for utilizing OWL as the representational language, and diverse approaches to constructing class hierarchies within the models. Noteworthy is the recurrent reuse of ontologies like OBO models (CIDO, GO, etc.) alongside CODO. The METHONTOLOGY approach emerges as a favored design methodology, often coupled with application-based or data-centric evaluation methods. Our study provides valuable insights for the scientific community and COVID-19 ontology developers, supplemented by comprehensive ontology metrics. By meticulously evaluating and documenting COVID-19 information-driven ontological models, this research offers a comparative cross-domain perspective, shedding light on knowledge representation variations. The present study significantly enhances understanding of CovOs, serving as a consolidated resource for comparative analysis and future development, while also pinpointing research gaps and domain emphases, thereby guiding the trajectory of future ontological advancements.

**Keywords:** COVID-19 ontologies, systematic literature review, ontological review, domain analysis, comparative study, ontology-driven models, knowledge representation.

## 1 Introduction

The emergence of the COVID-19 pandemic in December 2019 marked a pivotal and unprecedented moment in contemporary history, originating in Wuhan, China, and its rapid and relentless global spread resulting in significant loss of life, with the World Health Organization (WHO) reporting 769 million confirmed cases and 6.95 million documented deaths as of August 9, 2023 [1]. In a concerted response to this devastating crisis, a global immunization effort has administered an incredible number of 13.49 billion doses of vaccine by August 5, 2023, underscoring the collective resolve to mitigate the impact of the pandemic. As humanity grapples with this multi-faceted challenge, researchers from various disciplines and domains have come together to address various dimensions of the pandemic. This collaborative effort has led to an unprecedented influx of data and datasets, curated by governmental, non-governmental, and individual entities, underscoring an acute and urgent need for effective information management systems that can leverage and make sense of this vast amount of information. In this rapidly changing landscape, ontology-based systems have emerged as a beacon of hope, characterized by their semantic models and sophisticated data processing tools that promise to seamlessly integrate, analyze and visualize the complex web of data related to COVID-19. By providing the ability to glean insights from complex datasets, these systems provide a strategic advantage in decision-making processes, resource allocation, and disease prevention strategies. The potential of ontology-based systems to revolutionize the approach to pandemic management is undeniable. Against this backdrop, the present study embarks on a systematic exploration and evaluation of COVID-19 ontologies that have been proposed by the research community to address the multifaceted and ever-evolving challenges posed by the COVID-19 pandemic. Recognizing the vital role of ontologies as structured models capable of effectively representing information needs and formalizing complex processes, particularly in the field of medical knowledge representation and data sharing [2], this study seeks to unravel the intricacies of these ontology and their potential to reshape our understanding of the pandemic. The growing popularity of ontology-driven systems, as evidenced by their exponential growth in research [3], further highlights their critical importance in addressing the complex and multifaceted challenges that the COVID-19 pandemic continues to present. Building on notable existing studies, this study aspires to fill an important gap in the literature by undertaking a comprehensive comparative analysis that delves into key parameters such as scope, intended purpose, ontology granularity, modularity, formalism, vocabulary reuse, and domain coverage (discussed further in section 2 and section 3). In doing so, this study aims to provide a consolidated and indispensable resource for researchers, practitioners, and ontology developers who are actively engaged in the pursuit of effective ontology-driven COVID-19 information representation.

The main objective of this study is multifaceted. This involves the rigorous identification and comprehensive analysis of existing COVID-19 ontologies (CovOs) in the vast expanse of scientific literature. This comprehensive undertaking involves a detailed examination of their unique attributes, diverse design methodologies, and specific scope, complemented by an incisive examination of ontology granularity and coverage in the complex COVID-19 domain. Through these multi-faceted goals, this research strives to provide valuable insights that transcend disciplinary boundaries, serving the needs of the scientific community, ontology developers, and decision-makers. The contributions of this study are both profound and impactful. Beyond the extensive review of the CovOs literature, based on a concise list of parameters derived from [3], this study introduces a new parameter, "Ontology Coverage/Domain", which further enriches the structured analysis of the selected ontology. Furthermore, this study sheds light on potential research gaps in the area of COVID-19, identifying areas where additional ontologies may be needed to comprehensively address the complex landscape of the pandemic. This critical insight is poised to guide future research efforts, fostering a more holistic and informed approach to the development and deployment of ontology-based systems. The rest of the paper is organized as follows: section 2 discusses the related works in the literature and section 3 describes the methodology formulated for CovOs analysis. Section 4 unveils the findings of the study. Section 5 engages in a discourse on the discussion and limitations of the current study. Finally, section 5 provides a conclusive summary of the current study.

## 2 Related Work

Ontology, as discussed above, represents an explicit specification of a shared conceptualization. While fewer studies have formally reviewed ontologies created for capturing and reasoning COVID-19 information, it is imperative to conduct a comprehensive examination of existing works. This section delves into some of the prominent studies that have analyzed and assessed ontologies in this context.

Gao and Wang (2023) contribute an article [4] that delves into epidemic management data models from an ontological perspective, with a focus on enhancing data interoperability. The study evaluates and synthesizes various pertinent vocabularies and ontologies, including EPO [5], GeMInA [6], CIDO [7], IDO [8], CODO [9], COVIDCRFRAPID [10], and OPM [11]. Facets such as disease, person, organism, epidemiology, organization, medical personnel, medical activity, medical resource, infection transmission, statistics, and city are identified across these ontologies. While these existing models capture crucial aspects of epidemic scenarios, the study reveals persisting gaps, accentuating the necessity for further refinement and expansion. Bayoudhia et al. (2021) present an article [12] that offers an overview of biomedical ontologies for representing pandemics and infectious diseases. The authors emphasize ontologies' pivotal role in capturing and sharing knowledge related to various disease aspects, epidemiology, clinical features, and biology. The paper reviews ontologies developed for specific diseases, including malaria (IDOMAL) [13], dengue fever (IDODEN) [14], schistosomiasis (IDOSCHISTO) [15], COVID-19 Ontology [16], COVID-19 Surveillance Ontology [17], CODO, and CIDO. These ontologies leverage existing resources such as SNOMED CT, FOAF, and other ontologies like IDO Core and ChEBI. The authors discuss the use of tools like Protégé [18], DL reasoning, and SPARQL [19] queries for ontology development, evaluation, and utilization to advance disease understanding, control, and treatment endeavors. Ahmad et al. (2021) contribute an in-depth exploration of ontologies and tool support in the realm of COVID-19 analytics [20].

The study addresses challenges associated with the pandemic and emphasizes ontology-based solutions. Notable ontologies discussed encompass CODO, CIDO, IDO, and COVID-19 surveillance ontology. Yousefianzadeh et al. (2020) delve into the role of ontologies in managing the influx of COVID-19-related data in the medical domain [21]. The focus centers on ontologies stored in the BioPortal database, encompassing COVID-19-related ontologies such as CIDO, COVID-19 Ontology, IDO-COVID-19 [22], COVID-19 Surveillance Ontology, and CODO. These ontologies function as semantic tools to standardize and represent intricate and heterogeneous textual data linked to COVID-19. The article highlights the pivotal role of ontologies in supporting healthcare decision-making, data analysis, and knowledge dissemination during the COVID-19 pandemic. In the landscape of ontology evaluation, the notable studies mentioned above have yielded valuable insights, but a significant gap persists. As the field of COVID-19 research continues to rapidly evolve, a thorough study that encompasses a substantial amount of existing ontologies suitable for representing COVID-19 data remains conspicuously lacking. This gap is further exacerbated by the lack of comprehensive comparative analyzes that systematically explore critical parameters, including scope, intended purpose, ontology granularity, modularity, levels of formalism, vocabulary reuse, and extent of domain coverage. This obvious void in current scientific discourse underscores the fundamental motivation driving current research: to effectively fill this gap by performing systematic analysis, and facilitating a comprehensive comparison of CovOs. In undertaking this endeavor, our research aims to address the pressing need for a holistic and insightful assessment of the dynamic and evolving landscape of ontology-driven systems in the context of the COVID-19 pandemic. Table 1 provides a concise overview of the related studies.

**Table 1.** Overview of Related Studies on CovOs Evaluation and Comparative Analysis

| Study Authors | Main Focus | Ontologies/Vocabularies for Comparative Analysis | Key Contributions and Findings |
|---|---|---|---|
| Gao and Wang (2023) | Epidemic management data models | EPO, GeMInA, CIDO, IDO, CODO, COVIDCRFRAPID, OPM | Evaluated ontologies, identified recurring patterns, highlighted gaps. |
| Bayoudhia et al. (2021) | Biomedical ontologies for pandemics and infectious diseases | IDOMAL, IDOBRU, IDODEN, IDOSCHISTO, COVID-19, The COVID-19 Surveillance Ontology, CODO, and CIDO. | Reviewed disease ontologies, emphasized knowledge sharing, and utilized resources. |
| Ahmad et al. (2021) | Ontologies in COVID-19 analytics | CODO, CIDO, IDO, COVID-19 surveillance ontology. | Explored challenges, proposed solutions, and introduced reference architectures. |
| Yousefianzadeh et al. (2020) | Ontologies for managing COVID-19 data | CODO, COVID-19 Surveillance Ontology, CIDO, IDO-COVID-19, and Covid-19 ontology. | Stressed ontology-based tools, highlighted BioPortal ontologies' role. |

# 3 Methodology: CovOs Review and Analysis

The methodology employed for the review and analysis of CovOs is structured and systematic, drawing inspiration from [23] and tailored to the specific objectives of this study. The step-by-step approach is illustrated in Figure 1, followed by a comprehensive breakdown of each phase.

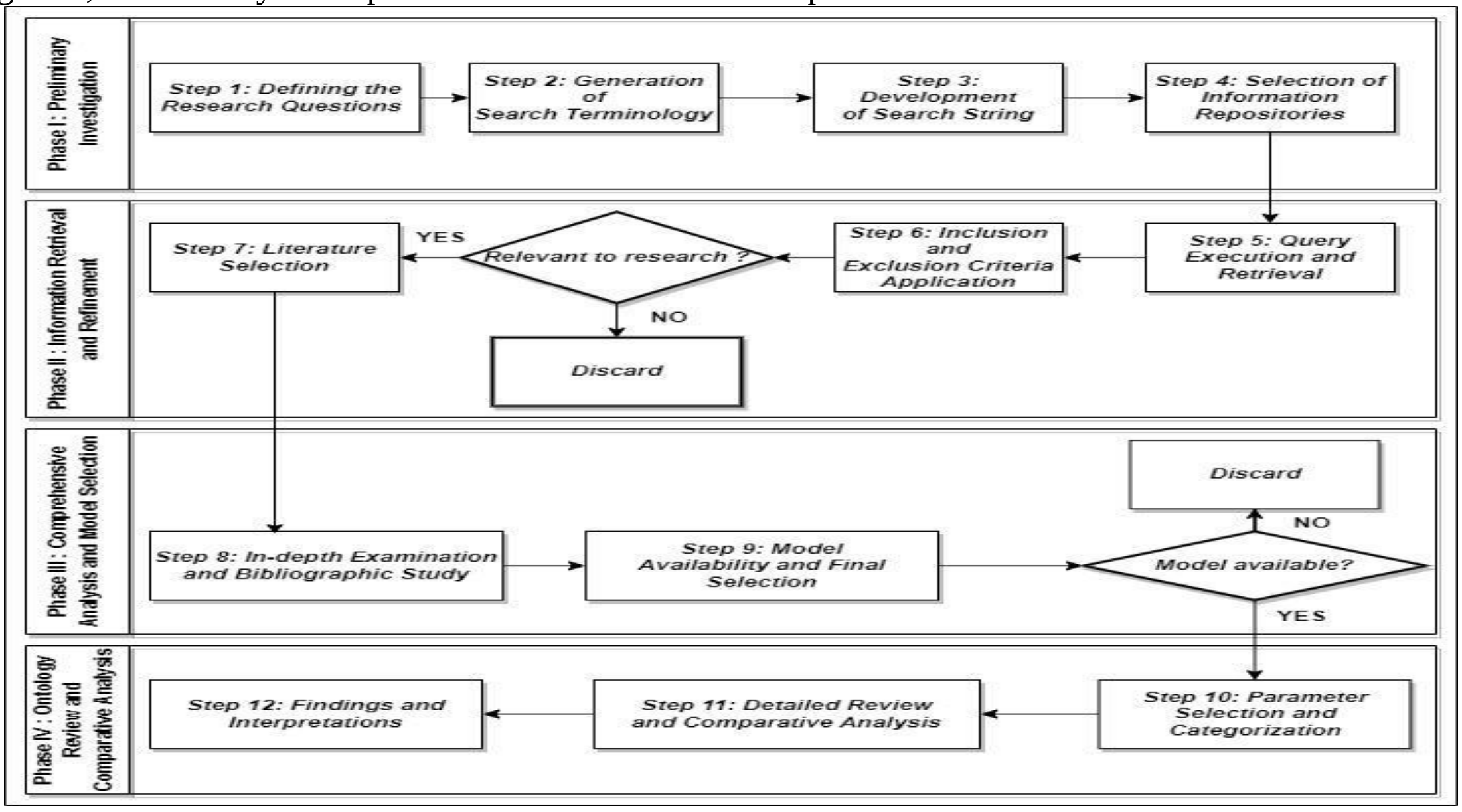

**Fig. 1**. The Methodology Workflow

**Phase I: Preliminary Investigation**

*Step 1: Defining the Research Questions*: The foundation of the methodology is laid by formulating pertinent research questions aligned with the study's objectives. These questions encompass the identification and analysis of existing ontological models for COVID-19-related information representation, along with an exploration of knowledge representation variations across domains. The framed research questions (RQ1-RQ5) are as follows: RQ1: What are the existing ontology-based models for representing COVID-19-related information? RQ2: How is COVID-19-related information represented using ontologies? RQ3: What are the domain coverages of existing COVID-19 ontologies? RQ4: What are the knowledge representation formalism languages used for creating COVID-19 ontologies? RQ5: How are the COVID-19 ontologies modeled with a focus on granularity, scope, modularity, level of formalism, and vocabulary (re)use?

*Step 2: Generation of Search Terminology:* Derived from the research questions, a set of pertinent search terms is constructed. These terms encapsulate the thematic essence of the research questions and aid in the subsequent literature search. The search terminologies (T1-T8) include: T1: COVID-19 ontology, T2: Coronavirus ontology, T3: COVID-19 knowledge representation, T4: COVID-19 model, T5: Epidemic ontology, T6: Infectious disease ontology, T7: Pandemic knowledge representation, T8: Disease ontology for COVID-19.

*Step 3: Development of Search String:* Search strings are formulated using combinations of the derived terminology (Step 2). Example search strings include S1: "ontology-based model for COVID-19", S2: "COVID-19" AND "ontology", S3: "coronavirus information" AND "ontology", S4: "ontology model for COVID-19", S5: "epidemic ontology" AND "COVID-19", S6: "infectious disease ontology" AND "COVID-19", S7: "pandemic knowledge representation", AND "COVID-19", S8: "disease ontology" AND "COVID-19".

*Step 4: Selection of Information Repositories:* Databases are chosen based on availability, reputation, and subject coverage. Databases include IEEE Xplore (https://ieeexplore.ieee.org/Xplore), Scopus (https://www.scopus.com/). ScienceDirect (https://www.sciencedirect.com/), Taylor and Francis (https://taylorandfrancis.com/), and Google Scholar (https://scholar.google.com/). Search strings are modified according to the databases.

**Phase II: Information Retrieval and Refinement**

*Step 5: Query Execution and Retrieval:* A comprehensive literature search is conducted using the formulated queries from Step 3 across the chosen information repositories. The search yields a pool of potential resources, which is subsequently refined, yielding 3437 documents. Duplicates are removed, resulting in 283 works.

*Step 6: Inclusion and Exclusion Criteria Application*: A systematic approach is employed to narrow down the resources based on inclusion and exclusion criteria, as detailed in Table 2. Criteria cover publication status, ontology description, and relevance to research questions, language, and function of ontology.

**Table 2**. Overview of Inclusion and Exclusion Criteria

| Categories | Inclusion Criteria | Exclusion Criteria |
|---|---|---|
| Publication status | Literature published in journals and conferences | Unpublished literature, non-journal sources |
| Description availability | Explicit ontology description | No ontology description |
| Relevance | Addresses research questions | Does not address research questions |
| Language | English language literature | Non-English language literature |
| Function of ontology | Models COVID-19 information | Non-COVID-19 information |

*Step 7: Literature Selection:* Applying the inclusion-exclusion criteria narrows down the literature to 38 works.

**Phase III: Comprehensive Analysis and Model Selection**

*Step 8: In-depth Examination and Bibliographic Study*: The selected literature is reviewed and analyzed, including a thorough examination of their bibliographies. This step aids in identifying relevant works that contribute to the research context.

*Step 9: Model Availability and Final Selection:* The focal point of this step is to assess the presence of CovOs in the selected literature. Works with available CovOs are prioritized for final inclusion, ensuring the suitability of the literature for subsequent comprehensive comparative analysis. Among the 38 works meeting inclusion criteria, 24 are selected based on CovOs availability and successful identification. This process entails extracting fundamental details (illustrated in Table 3) such as sponsored agencies, project names, ontology design patterns, utilized operations, illustrative classes, properties, and ontology metrics that characterize the ontologies. The summary of the chosen models is presented in Table 4, offering as guide for the upcoming analysis and discussions.

**Table 3.** Overview of Ontology selection parameters

| Criteria | Description |
|---|---|
| Sponsored Agencies | Identifies supporting agencies contributing to work development and promotion. |
| Project Name | Specifies the project name within which the study happens, particularly relevant for agencies overseeing multiple projects. |
| Ontology Design Pattern | Illustrates the structural framework of ontologies, categorized as either self-contained components (modular) or integrated parts of the overall ontology (non-modular). |
| Operation Used | Describes the methodology for ontology development, including operations combining disparate knowledge sources (integration), expanding existing ontology elements (extension), and removing unnecessary elements (pruning). |
| Example Classes and Properties | Highlights fundamental ontology components, encompassing concepts, data properties, and object properties, vital for accurate domain modeling. |
| Ontology Metrics | Offers insights into ontology structure with metrics such as class, subclass, property counts, and axioms, calculated even from partial ontology information. |

**Phase IV: Ontology Review and Comparative Analysis**

*Step 10: Parameter Selection and Categorization:* We extracted key comparison parameters from [3], organized them into categories, and introduced an additional parameter, "Ontology coverage/domain", to enhance the structured analysis of selected ontological models. Our study omits certain parameters like the "Focused phase" mentioned in [3] while incorporating "Ontology coverage/domain" to pinpoint potential research gaps and emphasize specific domains needing further exploration in comparative CovOs analysis. An outline of the crucial parameters for the comparative analysis can be found in Table 5.

*Step 11: Detailed Review and Comparative Analysis:* With the parameters defined (as outlined in Table 5), the subsequent phase involves a comprehensive evaluation of the selected ontologies. To facilitate this process, various spreadsheet tools can be utilized to organize the parameters systematically. In this study, data collection was executed using Microsoft Excel, and the collected information was structured within Table 6.

*Step 12: Findings and Interpretations:* Upon the completion of parameter measurement and tabulation, the focus shifts to in-depth analysis and uncovering significant findings. The assessment of the CovOs is succeeded by a detailed exploration of the insights and outcomes, subsequently discussed in depth (as elaborated in sections 4 and 5), offering insights into the intricacies and implications of COVID-19 information representation.

**Table 4.** Overview of selected ontologies

| Model No. | Model [Ref. No.] | Acronym | Sponsored Agencies | Project Name | Design Pattern | Operations Used | Example Classes and Properties | Ontology Metrics |
|---|---|---|---|---|---|---|---|---|
| M1 | Infectious Disease Ontology [24,25] | IDO | NF | NF | Modular | Integration, Extension | chronic infection, bacteria, pathogenic disposition, symptom, immune response, derives from, occurs in, participates in. | Class: 362 OP:43 DP:0 Individual 23 |
| M2 | Virus Infectious Disease Ontology [26] | VIDO | NIH / NLM | Biomedical Informatics and Data Science Research Training | Modular | Integration, Extension | Anatomical space, Symptom, blood, bodily fluid, host, virus, pathogen, disease, assay, replication, symptom, located in, occurs in, realized in. | Class 432 OP 32 DP 0 Individual 0 |
| M3 | International Classification of Diseases Ontology [27] | ICDO | University of Michigan. | Undergraduate Research Opportunity Program (UROP) | Modular | Integration, Extension | pathological derivation, pathological invasion, pathological transformation, bacterial pneumonia, viral pneumonia, bronchopneumonia, COVID-19 pneumonia, COVID-19, virus not identified, pneumonia due to SARS-associated coronavirus, respiratory failure, functionally related to, has icd10 code, has ICD10 name, has participant, has process profile, has prototype, has syndrome | Class 1313 OP 233 DP 1 Individual 4 |
| M4 | Coronavirus Infectious Disease Ontology [7] | CIDO | University of Michigan, University of North Dakota | NIH | Modular | Integration | pathological derivation, pathological invasion, pathological transformation, bacterial pneumonia, viral pneumonia, bronchopneumonia, COVID-19 pneumonia | Class 8775 OP 363 DP 18 Individual 457 |
| M5 | CIDO-COVID-19:Ontology of COVID-19 [28] | CIDO-COVID-19 | NIH | NF | Modular | Integration | diagnosis, personal information, diagnostic process, disease course, biological_process, viral process, gene, genome, diagnose, depends on, has participant, has concurrent part, is AA mutation of, is AA variant of, participates in, located in, may_prevent, admitted from, disposition of | Class 10386 OP 375 DP 25 Individual 458 |
| M6 | The COVID-19 Infectious Disease Ontology [8] | IDO-COVID-19 | NIH / NLM | NIH | Modular | Integration | Disorder, Infection, Protein, nucleic acid, pathogen, viral, assay, replication, symptoms, has part, has participant, located in, has quality, has functions | Class: 486 OP:43 DP:0 Individual 23 |
| M7 | COVID-19 surveillance Ontology [17] | COVID 19 | Public Health England / Wellcome | NF | Non-modular | Integration | Suspected COVID-19, Taking of swab for SARS-CoV-2 (severe acute respiratory syndrome coronavirus 2), Myocarditis caused by SARS-CoV-2 (severe acute respiratory syndrome coronavirus 2), possible COVID-19 | Class: 32 OP:0 DP:0 Individual:0 |
| M8 | COVID-19 Ontology [16] | COVID-19 | NF | NF | Modular | Integration | COVID virology, risk factor, transmission process, clinical aspect, laboratory finding, diagnosis, intervention, bodily process, transport of virus, isModelFor, isRiskFactorFor, isClinicalTrialFor, isVirologyFor | Class 2270 OP 9 DP 1 Individual 6 |
| M9 | Drugs for COVID-19 [29] | DRUGS4COVID19 | NF | DRUGS4COVID19 Spanish National Project | Modular | Integration | Active substance, Bibliographic resource, Chemical substance, Disease, Disorder, Drug, Effect, Paper, Paragraph, Sentence, Symptom, Caused by drug, Contains, Has active substance, Has effect, Has symptom, Is active substance of drug, Is symptom of disease, Is treated with drug, Date Disorder | Class 11 OP 17 DP 20 Individul 0 |
| M10 | WHO COVID-19 Rapid Version CRF Semantic Data Model [10] | COVIDCRFRAPID | NF | NF | Modular | Integration, Extension | data item, Questionnaire Subsection, Clinical or Research Assessment Questionnaire, Question_group, measurement unit label, datum label, Question, Boolean_Question, Demographics, Medication, Vital signs, Admission form, Question, is about, has part, part of, has measurement | Class 398<br>OP 6<br>DP 7<br>Individul 333 |
| M11 | Country Responses towards COVID-19 [30] | ROC: Ontology | German Federal Ministry of Education and Research | QURATOR | Modular | Integration | ResponseStatistics, CountryWiseStatistics, school_closing, contact_tracing, testing_policy, workplace_closing, facial_covering | Class 117<br>OP 87,DP 94<br>Individual 271 |
| M12 | Controlled Vocabulary for COVID-19 [31] | COVOC | HES-SO internal funding scheme, Covid-19 response program, #CuSToC grant | BICKL Research Project | Modular | Integration | host, disease, climate conditions, COVOC Term, Diagnosis, Chest Radiography, Vaccine, Clinical Data, confirmed case, contact tracing, Epidemiology, country, Hospitalization, weather changes, Drug Tolerance, genome, school closures, Protein, possible case, positive rate, lockdown, Cell Count, is disease model for, results_in, has_human_miRNA, | Class: 541<br>OP:179<br>DP:0<br>Individual:0 |
| M13 | Covid-19 Pandemic Ontology Model of Bogor City [32] | COPOMBOCY | NF | NF | Modular | Integration | Test Process, Agents, Organization, Person, Close Contact Tracking, Covid-19 Symptoms, Test Result, diseasetestedIn, docontactTracing, hasApplication, contactHistory, hasDiagnosis, hasContagion,hasDisease, hasResult, hasStatus, hasStatistics, travelHistory, hasDiagnosis, hasContagion, hasDisease, hasResult, hasStatus, hasStatistics, travelHistory | Class count 88<br>OP 29<br>DP 48<br>Individual count 229 |
| M14 | Homeostasis imbalance process ontology [33] | HOIP | NF | NF | Modular | Integration | ORF6 protein (SARS-CoV-2), senescence model, blood serum, cell cortex, abnormal cell tissue, Vaccine, bodily fluid, blood plasma, viral protein, spatiotemporal region, cytoplasmic pattern recognition receptor signaling pathway, aboral to, capable of part of, contained in, contains, expresses, finding of, has finding, has state, has participant, has part | Class: 12668<br>OP:180<br>DP:0<br>Individual:0 |
| M15 | The Coronavirus Disease Ontology [34] | CovidO | National Institute of Technology Kurukshetra | NF | Modular | Integration, pruning | Acute disease, Agent, Assembly, Bed, Birthday , City, CityWiseStatistics, Clinical finding (finding), CloseContact, admittingIn, belongsTo, causedBy, controlTo, diagnosisOf, address, admittedOn, closure , contactNo | Class: 116<br>OP:44<br>DP:63<br>Individual: 163 |
| M16 | COVID19 Ontology for analyzing the Karnataka Private Medical Establishments Data [35] | COKPME | NF | NF | Non-modular | Integration | Agent, Clinical Findings, Comorbidity, Diagnosis, Hospital, Location, hasDistrict, diahnosedFor, hasSymptom,hasPhone | Class: 25<br>OP:19<br>DP: 19<br>Individual: 41 |
| M17 | COVID-19 Common Datamodel [36] | GECCO | Internal Programs Fraunhofer vs Corona | COPERIMOplus | Non-modular | Integration | clinical trial, comorbidity, Complications Status, Patient Discharge, location of death, Migraine, supplemental oxygen therapy, Invasive blood pressure, ventilation frequency, TreatmentFor, has value domain, has unit, has data type, has Follow Up | Class: 5383<br>OP:14<br>DP: 0<br>Individual: 0 |
| M18 | Covid-19OntologyInPatternMedicine [37] | COVID-19-ONT-PM | NF | The Exploration Of Data Dynamics Of COVID-19 | Non-modular | Integration | Cell, Disease, FeatureOfObject, PathologyOfDisease, Patient, Protein, StructuralObject, SymptomOfDisease, Virus, Cell, Disease, FeatureOfObject, PathologyOfDisease, Patient, Protein, StructuralObject, SymptomOfDisease, Virus | Class: 299<br>OP: 11<br>DP: 1<br>Individual: 6 |
| M19 | Long Covid Phenotype Ontology [38] | LONGCOVID | Imperial College London | RECAP Predicting Risk of Hospital Admission in Patients with Suspected COVID-19 in a Community Setting study | Non-modular | Integration | COVID-19_-_test_OR_Diagnosis, No_further_analysis, Non-hospitalised_Long_COVID, Community_long_COVID, Long_COVID_diagnosis,_score,_OR_referral, Patients_/_Population_Included | Class 9<br>OP 0<br>DP 0<br>Individual 0 |
| M20 | VODANA-COVIDTERMS [39] | VODANACOVID | NF | NF | Non-modular | Extension | skos:Concept, skos:ConceptScheme, vodanacovidterms DateofPatientVisit, vodanacovidtermsAge, vodanacovidterms CityofCOVID19testcenter,vodanacovidtermsDateofCOVID19 test, vodanacovidtermsSerialNumber | Class 3<br>OP 5<br>DP 0<br>Individual 54 |
| M21 | ZonMW COVID-19 [40] | ZONMW-CONTENT | NF | NF | Non-modular | Integration | skos:Concept, skos:ConceptScheme | Class 3<br>OP 0<br>DP 0<br>Individual 273 |
| M22 | Ontology of SARS-CoV2 mutations [41] | SARSMUTONTO | NF | NF | Modular | Integration | mutation, SARS-CoV-2, genome, gene, non_structural_gene, structural_gene, variant, lineage, A.15, A.17, A.19, A.1, has_for_lineage, has_for_gene, has_for_alias, has_for_ description, has_for_WHO_name, label, mutation_name | Class 1616, OP 2<br>DP 6<br>Individual 2051 |
| M23 | OntoRepliCov [42] | ONTOREPLICOV | Agence Nationale de la Recherche (ANR) | PullCovAppart project | Modular | Integration | Genome ( i.e., Accessory_protein, Non_Structural_Protein, Polyprotein, UTR5), Replication_element (i.e., Endoribonuclease, ER, nsp12, RNA_Negative_Messenger, RNA_Positive, Subgenomic_Positive, tRNA), has_Rank, status, has_ending_base, has_first_base, has_first_codon | Class 85<br>OP 13<br>DP 4<br>Individual 15 |
| M24 | The COviD-19 Ontology for cases and patient information [9] | CODO | Indian Statistical Institute | Integrated and Unified Data Model for Publication and Sharing of prolonged pandemic data as FAIR Semantic Data: COVID-19 as a case study | Modular | Integration | Agent, Clinical finding (finding), Comorbidity, Diagnosis, Disease, Exposure to COVID-19, Gender type, Place, Statistics, Status, Symptom, Test result, TestData, admitted in, contracted virus from, hadCovidTest, has diagnosis, is covid-19 statistics of, age, date | Class 90<br>OP 73<br>DP 50<br>Individual 271 |

**Abbreviations used:** NIH- National Institutes of Health, NLM- National Library of Medicine, NF- Not Found, OP- Object Property, DP- Data Property.

**Table 5.** Overview of key parameters for the comparative analysis

| Parameters | Description |
|---|---|
| Purpose | Represents the intended objective of the ontology, detailing its purpose and the specific information domain it aims to cover. |
| Ontology Type | Classifies the ontology into distinct types, such as General Ontology, Domain-specific Ontology, Application-specific ontology to identify its focus and scope within the broader context. Where, General Ontology means a comprehensive framework of fundamental concepts and relationships that spans multiple domains. Domain Ontology is a structured representation of concepts and relationships within a specific subject area or field, and Application-Specific Ontology is a specialized knowledge representation tailored to a particular context or software application. |
| Ontologies/Vocabularies Reused | Identifies and lists existing ontologies or vocabularies incorporated in the ontology's construction, highlighting resource reuse for knowledge representation. |
| Design Methodology | Describes the systematic approach and principles followed during the creation of the ontology, ensuring adherence to established ontological methodologies. |
| Class Hierarchy Development | Explains the method by which the hierarchy of classes within the ontology is developed, whether through top-down, bottom-up, combination or other systematic approaches. Methods for constructing class hierarchies, involve either starting from general concepts and refining (top-down), or from specific entities and aggregating (bottom-up), or a blend of both approaches. |
| Representation Language | Specifies the language employed for the representation of the ontology. |
| Level of Formality | Indicates the level of formality in the ontology's language expressivity, categorizing it as informal, formal, or semi-formal, determining its machine-process ability. The degree of structure in representing information, ranging from casual and unstructured (informal) to structure with defined rules and syntax (formal), with an intermediate state (semi-formal). |
| Ontology Editor | Refers to the software tool or platform utilized for creating, visualizing, and modifying the ontology, enhancing its development and maintenance capabilities. |
| Evaluation | Details the methods and processes used to evaluate the ontology's effectiveness, validation, and trustworthiness in capturing and representing domain knowledge. |
| Ontology Coverage/Domain | Specifies the specific domains or areas of information that the ontology is designed to cover, outlining its scope within the broader knowledge landscape. |

## 4 Findings

In Table 6, we present an overview of the selected twenty-four CovOs (M1 to M24). The table demonstrates the varied purposes of CovOs (M1 to M24), spanning from deepening insights into virus-related domains and the biomedical facets of COVID-19, to supporting primary care surveillance, elucidating drug associations, facilitating data integration, and providing comprehensive data representation. These models aim to provide structured knowledge on infectious illnesses, expand existing ontologies, aid medical decision-making, standardize terminology, annotate literature, depict COVID-19 progression, characterize genetic aspects, and enable data publishing. By addressing different aspects of COVID-19 and related domains, these CovOs contribute to improved research, analysis, and decision support in the fight against the pandemic. The ontology types of the overall models vary based on their intended purposes and applications. For instance, models M1 to M3 serve as general ontologies, providing foundational concepts for infectious illnesses, virus-specific terms, and structured biomedical representations of disease classifications. Models M4 to M9 are domain ontologies, focusing on comprehensive biomedical representations of coronavirus infectious disease, drug associations, and patient data for improved medical insights. Models M10 and M11 are application ontologies, designed to structure patient data and analyze global COVID-19 responses, while M12 acts as an application ontology for biomedical literature annotation. Models M13 to M24 fall under the category of domain ontologies, capturing various aspects of the COVID-19 pandemic, such as regional knowledge, virus progression, genome characterization, and replication processes. These domain ontologies contribute to a comprehensive understanding of COVID-19 and support a range of research and practical applications. The ontology vocabulary reused across the different models provides a foundation for their knowledge representation and interoperability. M1 and M2 utilize OBO and DC. In M3, FOAF, OBO, BFO, OGMS, UBERON, and PATO contribute to structured biomedical representations. M4 and M5 reuse OBO models, FOAF, and SNOMED CT for enhanced COVID-19 understanding. M6 leverages OBO and DC, while for M7 we have not found any mentioned methodology. M8 integrates OBO, FOAF, and SKOS for enriched domain knowledge, and M9 adopts DC term and ATC for bibliographic resources. M10 involves DC, OBO models, and GEO-Ont. M11 benefits from CODO, SNOMED, FOAF, and SCHEMA. M12 incorporates

OBO models. M13 and M14 rely on CODO and CIDO, while M14 further integrates Basic Formal Ontology, and various OBO models. M15 utilizes Schema, OBO, CODO, FOAF, SCHEMA, and IBO. M16 and M17 integrate Schema, FOAF, CODO, SNOMED CT, and OBO. M18 involves CIDO, COVID-19 Ontology, and GO. M19 and M22 are unspecified (NF). M20 and M21 leverage SKOS for structured vocabularies. M23 adopts OBO, and M24 combines FOAF, ORD, SCHEMA, SNOMED CT, and OBO for comprehensive data representation. *The design methodologies* employed across these models guide the systematic creation of ontologies, ensuring their effectiveness and relevance. Models M1, M2, M4, and M5 follow the OBO methodological principles, promoting consistency and compatibility within biological and biomedical ontologies. M3 adopts the eXtensible Ontology Development (XOD) strategy for flexible ontology creation, while M7 utilizes METHONTOLOGY to establish surveillance systems. M9 incorporates temporal information using the LOT approach. M11 employs the NeOn methodology for systematic ontology construction, and M13, M14, and M15 utilize METHONTOLOGY and Diligent approaches to enhance knowledge representation. M16, M17, and M24 continue with METHONTOLOGY and YAMO methods to facilitate comprehensive data analysis and publication. Some models are not specified (NF), and each methodology contributes to the ontologies' structure, usability, and relevance. *Class hierarchy development* varies across models: M1-M6, M8, M9, M12, M14, M15 M18, M22, and M23 follow top-down; M7, M10, M13, M16, M17, M19-M21, and M24 adopt bottom-up; and M11 use combination approaches. *Representation language* signifies the medium for depicting concepts and relationships, such as OWL and RDFS. This parameter highlights the prevalent language choice for representation. Our analysis indicates that OWL was predominantly used in the ontologies developed, as demonstrated in Table 6. *Level of Formality* pertains to the extent of rigor in ontology development, such as Semi-formal for M1-M6, M13-M15 and Informal for M7-M12, M16-M23 models. M24 adopts a Formal approach, showcasing the diverse formalities employed in constructing the ontologies. The *Ontology Editor* used varies across the models, with Protégé being the predominant choice for M1-M23 and M24, highlighting its widespread adoption. M6 employs an unspecified editor, while M20 and M21 utilize the sheet2rdf GitHub workflow. *Ontology Evaluation methods* vary across models. Evaluation serves as a foundation for designing new ontologies and enables updates when necessary. Various evaluation approaches exist, including comparing with golden standards, application-based assessment, data-based comparisons, expert or human evaluations, and task-based and criteria-based evaluations. Depending on factors like standard ontology availability, expertise, data, and application, relevant evaluation methods are chosen. M1 and M2 employ HermiT, Pellet reasoner, Mace4 model checker, and Prover9. M3 uses Hermit reasoner, user feedback, and application-based evaluation. M4 and M5 rely on HermiT, Pellet reasoner, Mace4 model checker, and Prover9. M6 applies application-specific validation, while M7 involves expert evaluation and external consensus exercise. M9 uses OOPS! and data-driven evaluation. M11 and M12 utilize SPARQL and Lucene Elasticsearch engine metadata parsing, respectively. M13 ensures consistency through Pellet reasoner and expert/non-expert validation. M14 employs Hermit27, ELK28, and SPARQL. M15 uses SPARQL query and OOPS! Pitfall Scanner. M16, M17, and M23 utilize SPARQL queries. M24 combines reasoner-based and SPARQL query evaluations. M8, M10, M18, M19, M20, M21, M22, and M23 are not found (NF). The *ontology coverage/domain* defines the extent of the ontologies, encompassing a range of disease-related dimensions. The ontology coverage or domain defines the breadth of ontological representation across various disease-related dimensions. Among the 24 CovOs (M1-M24) and the identified 23 distinct domains (D1-D23), each model is associated with specific domain numbers based on its coverage. For instance, Models M1, M2, and M5 encompass domains such as etiology, epidemiology, pathogenesis, and virology. Models M6, M7, and M8 emphasize diagnosis, prevention, and therapy. Model M14 is tailored toward procedures and resources, while M15 and M16 delve into immunology and ethics respectively. M17 explores demographics, race, and ethnicity, and M20 investigates the influence of weather. Models M21 and M22 pertain to laboratory tests and locations, while M23 is primarily centered on statistics and data analysis. Lastly, Model M24 spans domains including etiology, epidemiology, pathogenesis, virology, diagnosis, prevention, therapy, procedures, resources, immunology, ethics, demographics, race and ethnicity, weather influence, laboratory tests, locations, and statistics, and data analysis.

This extensive range of disease-related domains encompassed by these ontologies fosters a comprehensive understanding and analysis of various aspects pertaining to disease and health.

# 5 Discussion and Limitations

## 5.1 Comprehensive Exploration and Insights

This section presents a comprehensive exploration of the research questions posed in section 3 (RQ1-RQ5) in the context of the 24 CovOs (M1 to M24). Collectively, these models offer a panoramic insight into the landscape of ontology-based approaches for representing COVID-19-related information, showcasing the diversity in ontological representations during the pandemic (RQ1). The analysis delves into the methodologies, vocabularies, and representation languages employed to depict COVID-19-related information, providing a detailed understanding of how ontologies capture this intricate domain (RQ2). Moreover, the models reveal an expansive domain coverage, spanning various disease-related aspects such as etiology, epidemiology, virology, primary care surveillance, government responses, and regional contexts, effectively addressing a wide spectrum of knowledge domains (RQ3). For example, models like M1, M2, and M5 span domains such as etiology, epidemiology, pathogenesis, and virology. M7 focuses on primary care surveillance, M11 on global government responses, and M13 on a specific regional context. Other models delve into immunology, genetics, patient data, and clinical aspects. Thus, the ontology models comprehensively cover various domains related to COVID-19. Exploring the world of knowledge representation formalism within the ontology models reveals a prominent pattern – the widespread utilization of the OWL language. This formalism preference for OWL is evident in several models, including M1, M5, M13, M15, M16, M17, and M24. The prevalence of OWL across these models highlights its significant contribution to shaping the landscape of ontology-based COVID-19 modeling (RQ4). The attributes of the models, encompassing granularity, scope, modularity, formalism, vocabulary reuse, and domain coverage, are examined, providing a comprehensive view of their characteristics and contributions (RQ5). Models like M24 exhibit a high level of formalism, while others, such as M6 and M7, adopt more informal approaches. Granularity varies across models, with some focusing on specific aspects like virus progression (M13) or clinical findings (M14). Vocabulary reuse is a common practice, promoting interoperability. As a whole, these CovOs, spanning from M1 to M24, collectively fulfill the outlined research questions, enhancing their role in advancing research, analysis, and decision-making in the pandemic landscape.

**Table 6.** Overview of selected CovOs

| Model No. | Purpose | Ontology Type | Ontologies/ Vocabularies reused | Design Methodology | Class hierarchy development | Representation Language | Level of Formality | Ontology Editor | Evaluation | Ontology Coverage/ Domain |
|---|---|---|---|---|---|---|---|---|---|---|
| M1 | To provide an interoperable ontology for infectious illnesses, incorporating clinical and biological details, while building upon the foundational 'disease' entity from the Ontology for General Medical Science (OGMS). | General Ontology | OBO (BFO, OGMS, NCBITaxon, UBERON, IAO, GO, VSMO etc), DC, | OBO methodological principles | Top-down | OWL | Semi-formal | Protégé | HermiT, Pellet reasoner, Mace4 model checker, Prover9 | D1, D2, D4, D5, D6, D8, D9, D10, D14, D17, D19. |
| M2 | To expand IDO by including virus-specific concepts according to OBO Foundry guidelines, comprehensively covering virology-related terms, and integrating information from OBO Foundry ontologies, thereby enhancing our understanding of virus-related domains. | General Ontology | OBO (BFO, CHEBI, OBI, CL, OMRSE, OGMS, NCBITaxon, UBERON, IAO, GO, VSMO, etc), DC, CCO | OBO methodological principles | Top-down | OWL | Semi-formal | Protégé | HermiT, Pellet reasoner, Mace4 model checker, Prover9 | D1, D2, D4, D5, D6, D8, D9, D10, D17. |
| M3 | To offer a structured and coherent biomedical ontology, facilitating the logical representation of terms and relationships pertaining to the International Classification of Diseases (ICD). | General Ontology | FOAF, OBO, BFO, OGMS, UBERON, PATO | eXtensible Ontology Development (XOD) strategy | Top-down | OWL | Semi-formal | Protégé | Hermit reasoner, Applications, user feedback | D1, D4, D5, D6, D7, D8, D17. |
| M4 | To establish a comprehensive, community-driven biomedical ontology for coronavirus infectious disease, encompassing coronavirus-specific entities, concepts from related ontologies, and open-source collaboration. | Domain | OBO (i.e., CIDO, OGMS, IDO, TRANS, SYMP, DRON, NDF-RT, CHEBI, NCIT, OAE, PW, VO, SO, PR), FOAF, SNOMED CT | OBO methodological principles | Top-down | OWL | Semi-formal | Protégé | HermiT, Pellet reasoner, Mace4 model checker, Prover9 | D1, D2, D4, D5, D6, D8, D9, D10, D17. |
| M5 | CIDO-COVID-19 enhances CIDO by offering a comprehensive COVID-19 ontology, aiding medical professionals and researchers with improved concepts, diagnosis, and treatment information while adhering to OBO and BFO guidelines. | Domain | OBO (i.e., CIDO, OGMS, IDO, TRANS, SYMP, DRON, NDF-RT, CHEBI, NCIT, OAE, PW, VO, SO, PR), FOAF, SNOMED CT | OBO methodological principles | Top-down | OWL | Semi-formal | Protégé | HermiT, Pellet reasoner, Mace4 model checker, Prover9 | D1, D2, D4, D5, D6, D8, D9, D10, D14, D15, D16, D17. |
| M6 | IDO-COVID-19 extends IDO and VIDO, focusing on SARS-CoV-2 infection and COVID-19, aligning with OBO Foundry standards and cross-products of Foundry ontologies, to comprehensively cover epidemiology, classification, pathogenesis, and treatment. | Domain | OBO, DC | OBO methodological principles | Top-down | OWL | Semi-formal | NF | Application specific validation. | D1, D2, D5, D6, D8, D9, D10, D14, D15, D16, D17. |
| M7 | To enable primary care SARS-CoV-2 surveillance by integrating data from diverse medical record systems to monitor COVID-19 cases and related respiratory conditions. | Application | NF | METHONTOLOGY | Bottom-up | OWL | Informal | Protégé | Expert evaluation, External validation using a rapid Delphi consensus exercise | D1, D2, D6, D8, D9, D10, D14, D15, D16, D17. |
| M8 | To represent molecular interactions, medical concepts, and epidemiological aspects of COVID-19, featuring a scalable SARS-CoV-2 entity and an extensive range of chemical entities for drug repurposing, while its efficacy is evaluated through analysis of Medline and the Allen Institut''s COVID-19 corpus. | Domain | OBO, FOAF, SKOS | NF | Top-down | OWL | Informal | Protégé | NF | D1, D2, D4, D5, D6, D8, D9, D10, D11, D14, D15, D16, D17, D20, D21, D22, D23. |
| M9 | To categorize drug-COVID-19 associations, encompassing core concepts of drug effects, diseases, symptoms, and chemical substances, leveraging coronavirus-related literature to address COVID-19 challenges. | Domain | DC term ( for bibliographic resources such as abstract, title, license, and source), ATC | LOT | Top-down | OWL | Informal | Protege | OOPS! And Data driven evaluation | D1, D4, D5, D8, D9, D10, D13, D16, D17. |
| M10 | To offer semantic references to the WHO's COVID-19 RAPID case record form, enabling structured representation of patient data for various applications, including graph-based machine learning. | Application | DC, OBO (IAO, NCIT, BFO,), GEO-Ont | NF | Bottom-up | OWL | Informal | Protégé | NF | D1, D2, D7, D8, D15, D16, D17. |
| M11 | To facilitate data integration from diverse sources to analyze the effectiveness and side effects of COVID-19 government responses globally, utilizing an ontology for assessing and interlinking national measures, and supporting insightful inquiries from the data. | Application | CODO, SNOMED, FOAF, SCHEMA | NeOn, METHONTOLOGY | Combination | OWL | Informal | Protégé | SPARQL | D1, D3, D8, D11, D12, D13, D16, D17, D18, D19, D21, D22, D23. |
| M12 | To facilitate seamless biomedical literature annotation for core databases and ELIXIR tools, using OBO ontologies to enhance connections to resources like the COVID-19 Data Portal. | Application | OBO (CHEBI, CHMO, EFO, NCIT, OMIT, PR, PATO, etc.) | NF | Top-down | OWL | Informal | Protégé | Lucene Elastic search engine metadata parsing, | D1, D2, D3, D5, D6, D7, D9, D10, D11, D12, D13, D14, D15, D16, D17, D18, D19, D20, D21, D22, D23. |
| M13 | To offer comprehensive knowledge on the COVID-19 pandemic in Bogor City, Indonesia, encompassing information about agents, transmission, diagnosis, location, restrictions, statistics, and test-related aspects. | Domain | CODO CIDO | METHONTOLOGY | Bottom-up | OWL | Semi-Formal | Protégé | Consistency testing by Pellet reasoner, Representational testing by expert and non-expert. | D1, D3, D8, D11, D12, D13, D15, D16, D17, D18, D19, D21, D22, D23. |
| M14 | HOIP ontology systematically models COVID-19 progression by analyzing homeostatic imbalances, aligning design patterns with CIDO, and organizing terms into a three-layer structure, encompassing general, biomedical, and imbalance-dependent concepts, thus enhancing our comprehension of disease mechanisms. | Domain | Basic Formal Ontology, VO, UBERON, Cell Ontology, NCBI Taxon, ChEBI, GO, PATO, OGG, INOH, HINO, DOID, HP, CIDO, BSPO, WBbt. | Diligent and METHONTOLOGY | Top-down | OWL | Semi-Formal | Protégé | Hermit, ELK , SPARQL | D1, D8, D9, D15, D16, D17. |
| M15 | To provide a comprehensive, structured metadata model for multi-modal coronavirus data, covering diverse aspects and dimensions of COVID-19, enabling applications to implement a consistent and interoperable metadata vocabulary. | Domain | Schema,OBO, CODO, FOAF, SCHEMA, IBO | Diligent and METHONTOLOGY | Top-down | OWL | Semi-Formal | Protégé | SPARQL query and OOPS! Pitfall Scanner. | D1, D2, D4, D5, D6, D8, D9, D10, D11, D14, D15, D16, D17, D18, D19, D20, D21, D22, D23. |
| M16 | To analyze pandemic data by capturing attributes from private clinics in Karnataka, encompassing patient details, clinical findings, diagnosis, and hospital information to assess the pandemic situation. | Domain | Schema, FOAF, CODO, SNOMED CT and OBO. | METHONTOLOGY | Bottom-up | OWL | Informal | Protégé | SPARQL query | D1, D2, D4, D5, D6, D8, D9, D10, D15, D16, D17, D18, D19, D21, D22, D23. |
| M17 | To establish interoperability and compatibility among diverse COVID-19 datasets by constructing a Common Data Model (CDM) with standardized variables, encompassing patient information, disease-specific data, medication, risk factors, diagnostics, demographics, findings, and more, facilitating comprehensive analysis and research. | Domain | Schema, FOAF , SNOMED CT and OBO | METHONTOLOGY | Bottom-up | OWL | Informal | Protégé | SPARQL query | D2, D4, D5, D6, D8, D9, D10, D11, D14, D15, D16, D17, D23. |
| M18 | To enhance pandemic response by focusing on scientific findings, enabling integrated and systematic medical decision-making through Pattern Medicine (PM) and Generalized Biomedical Dynamics (GBMD), specifically addressing relationships between molecular and clinical levels for improved understanding and decision support. | Domain | CIDO, COVID-19 Ontology, GO, | NF | Top-down | OWL | Informal | Protégé | NF | D1, D2, D3, D4, D5, D6, D9, D10, D11, D13, D15, D16, D17, D18, D19, D21. |
| M19 | To offer insights into primary care presentations of prolonged COVID-19 symptoms beyond 4 weeks, encompassing various diagnoses and clinical definitions, aiding understanding and implications of this condition. | Domain | NF | NF | Bottom-up | OWL | Informal | Protégé | NF | D1, D4, D5, D8, D9, D12, D15, D16, D17, D18, D19. |
| M20 | To standardize controlled terms and semantic properties within the COVID-19 generic template of VODANA, facilitating consistent data representation across facilities and encompassing elements such as cause of death, COVID-19 results, vaccination status, and health metrics. | Domain | SKOS | NF | Bottom-up | OWL | Informal | sheet2rdf GitHub workflow | NF | D5, D9, D10, D12, D13, D14, D15, D16, D17, D18, D19, D21, D23. |
| M21 | To compile terms for the ZonMW COVID19 program, utilizing vocabularies to describe project and data content metadata, focusing on essential aspects including patients, proteins, viruses, hosts, cells, organs, pathology, diseases, and symptoms. | Domain | SKOS | NF | Bottom-up | OWL | Informal | sheet2rdf GitHub workflow | NF | D2, D5, D6, D8, D9, D13, D14, D15, D16, D17. |
| M22 | To characterize the SARS-CoV2 genome, encompassing its genes, mutants, and mutations in a hierarchical manner, providing detailed information about their lineage, appearance, aliases, descriptions, and WHO names, contributing to a comprehensive understanding of the virus's genetic structure. | Domain | NF | NF | Top-down | OWL | Informal | Protégé | NF | D6, D7, D13, D14. |
| M23 | To depict distinct phases of SARS-CoV-2 replication, detailing genome components and replication elements aiding comprehension of the virus's replication process through attributes like rank, status, and base information. | Domain | OBO | NF | Top-down | OWL | Informal | Protégé | SPARQL query | D6, D7, D13, D14. |
| M24 | Serves as a knowledge graph model to publish COVID-19 data on the web, focusing on cases across locations, patient data (symptoms, travel history), clinical findings, diagnosis, transmission, and resource requirements, facilitating comprehensive data representation and analysis. | Domain | FOAF, ORG, SCHEMA, SNOMED CT, OBO | YAMO | Bottom-up | OWL Domain Specific | Formal | Protégé | Reasoner based and SPARQL query | D1, D2, D4, D5, D6, D8, D9, D10, D11, D14, D15, D16, D17, D18, D19, D20, D21, D22, D23. |

**Abbreviations used :** OBO - Open Biological and Biomedical Ontologies, BFO - Basic Formal Ontology, OGMS - Ontology of General Medical Science, NCBITaxon - National Center for Biotechnology Information Taxonomy, UBERON - Integrated Cross-Species Anatomy Ontology, IAO - Information Artifact Ontology, GO - Gene Ontology, VSMO - Vaccine Safety Markers Ontology, DC - Dublin Core, CHEBI - Chemical Entities of Biological Interest, OBI - Ontology for Biomedical Investigations, CL - Cell Ontology, OMRSE - Ontology of Medically Related Social Entities, FOAF - Friend of a Friend, PATO - Phenotypic Quality Ontology, CIDO - Coronavirus Infectious Disease Ontology, IDO - Infectious Disease Ontology, TRANS - Translational Medicine Ontology, SYMP - Symptom Ontology, DRON - Disease Ontology, NDF-RT - National Drug File Reference Terminology, NCIT - National Cancer Institute Thesaurus, OAE - Ontology of Adverse Events, PW - Plant Trait Ontology, VO - Vaccine Ontology, SO - Sequence Ontology, PR - Protein Ontology, GEO-Ont - Gene Expression Omnibus Ontology, SNOMED - Systematized Nomenclature of Medicine, SCHEMA - Schema.org, ATC - Anatomical Therapeutic Chemical Classification System, SKOS - Simple Knowledge Organization System, ORD - Ontology of Regulatory Elements, OWL - Web Ontology Language, SPARQL - SPARQL Protocol and RDF Query Language, NF - Not Found.

**Ontology Coverage/ Domain D1 to D23 defines-**

D1 - Etiology: Study of the causes and origins of diseases or conditions., D2 - Epidemiology: Study of the patterns, causes, and effects of health and disease conditions within populations., D3 - Ethnology: Analysis of the cultural, social, and behavioral factors that influence health and disease within specific ethnic groups., D4 - Transmission: Study of how diseases or infections are spread between individuals., D5 - Pathogenesis: Investigation of the mechanisms and processes that lead to the development of diseases., D6 - Virology: Study of viruses, their properties, classification, and interactions with their host organisms., D7 - Genomics: Exploration of the complete set of genes within an organism, including their functions and interactions., D8 - Diagnosis: Identification of diseases or conditions based on signs, symptoms, and medical tests., D9 - Prevention: Strategies and measures to reduce the risk of diseases or conditions from occurring., D10 - Therapy/Medication: Approaches and drugs used for the treatment of diseases or conditions. D12 - Preparation: Steps taken to get ready for disease outbreaks or emergencies., D11 - Therapeutic Safety: This domain encompasses the study of risk factors, adverse effects, and drug interactions associated with COVID-19 therapeutic interventions, focusing on ensuring the safety and well-being of patients undergoing treatment.D13 - Control measures: Strategies and actions to manage and mitigate the spread of diseases., D14 - Procedure: Specific medical or research methods used in disease management or study., D15 - Immunology: Study of the immune system and its response to infections and diseases., D16 - Ethics: Examination of moral and ethical considerations in healthcare, research, and disease management., D17 - Phenomenon: Observations and occurrences related to diseases, their impacts, and effects., D18 - Demographics: Statistical study of populations, including their size, structure, and characteristics., D19 - Race and Ethnicity: Analysis of how different racial and ethnic groups are affected by diseases and healthcare., D20 - Weather: Exploration of how climatic conditions influence the spread and impact of diseases., D21 - Laboratory Test: Medical tests and procedures performed in laboratories for disease diagnosis and monitoring., D22 - Location: Geographic areas and regions where diseases occur and are studied., D23 - Statistics and Data: Collection, analysis, and interpretation of data to understand disease patterns and trends.

## 5.2 Identifying Research Gaps and Domain Emphasis

In addition to the insights gained from our analysis, the identification of research gaps and domain emphasis underscores areas of potential focus for future ontological development. Among the examined CovOs, certain domains, notably D1 (Etiology), D2 (Epidemiology), D4 (Transmission), D5 (Pathogenesis), D6 (Virology), D8 (Diagnosis), D9 (Prevention), and D17 (Phenomenon), emerge as well-represented across multiple models, signifying their significance in the ontological landscape. Conversely, domains such as D3 (Ethnology), D7 (Genomics), D11 (Therapeutic Safety), D13 (Control Measures), D19 (Race and Ethnicity), D20 (Weather), D21 (Laboratory Test), D22 (Location), and D23 (Statistics and Data) exhibit limited coverage, suggesting potential research gaps in these areas. This observation highlights opportunities for extended research efforts to enhance the depth and breadth of ontological representation within these domains. Furthermore, it is noteworthy that while some of the CovOs adopt formal methodologies, a majority of them lean towards informal or semi-formal approaches, indicating a potential avenue for further advancement in terms of formal ontology modeling. Thus, the analyzed CovOs not only aid in recognizing well-addressed domains but also serve as guides for directing attention toward domains that would benefit from enhanced ontological focus, thereby offering valuable guidance for shaping future research endeavors.

### 5.3 Limitations and Scope

It is important to acknowledge certain limitations inherent to this study. The inclusion criteria for the selection of CovOs is guided by specific research objectives, potentially resulting in the omission of other pertinent models. Moreover, our analysis is confined to a comprehensive assessment of existing CovOs based on the specified parameters outlined in Section 3. The scope of this research is centered on the representation of COVID-19-related information within ontologies, and as such, the analysis may not encompass all facets or variables within the broader domain of ontological research. Despite these limitations, the study provides valuable insights into the domain coverage, methodologies, and characteristics of CovOs relevant to COVID-19, thereby assisting researchers in advancing comprehension and effective utilization of ontologies within the pandemic context.

## 6 Conclusion

This study has undertaken a comprehensive exploration of ontology-based approaches for representing COVID-19-related information, offering valuable insights into the diverse landscape of ontological representations. Through the analysis, the study has not only addressed the framed research questions but has also contributed to the advancement of knowledge in several key aspects. The study has successfully identified and examined twenty-four CovOs (M1 to M24) spanning a wide range of domains, from etiology and epidemiology to virology, diagnosis, prevention, and more. This extensive coverage has provided an aerial perspective of the ontological landscape, fulfilling the contribution statement of this work. Furthermore, the study's detailed analysis of methodologies, representation languages, class hierarchies, and ontology evaluation methods has enriched our understanding of the characteristics and attributes of these ontological models. By comparing and contrasting the different models, the study has highlighted both distinct features and commonalities, thereby offering insights into best practices and potential areas for improvement. In fulfilling its contribution statement, this study has not only provided a comprehensive overview of CovOs but has also identified potential gaps in research. The emphasis on certain domains across several models, in contrast to the limited representation of others, highlights the need for additional ontological development in specific areas. This work lays the groundwork for future research endeavors by guiding researchers toward domains that require additional focus and refinement. Building upon the insights gained from this study, future research endeavors could delve into several promising directions. Firstly, an extended exploration of the identified research gaps and unaddressed domains within existing COVID-19 ontologies (CovOs) could provide a foundation for developing specialized ontologies to fill these gaps and enhance the comprehensiveness of COVID-19 data representation. Furthermore, the refinement and expansion of the parameter set used for ontology comparison, potentially incorporating emerging criteria and advanced evaluation methodologies, could contribute to a more nuanced and comprehensive analysis of CovOs. As the pandemic landscape continues to evolve, ontologies will play a crucial role in advancing research, analysis, and decision-making, and this study provides a solid foundation for researchers to build upon further to enhance our understanding of COVID-19 and its multifaceted dimensions.